\documentstyle[prb,aps,12pt]{revtex}

\begin{document}
\title{Ground state features of the Fr\"ohlich model}
\author{G. De Filippis, V. Cataudella, V. Marigliano Ramaglia, 
C.A. Perroni, and D. Bercioux}
\address{Coherentia-INFM  and Dipartimento di Scienze Fisiche, \\
Universit\`{a} degli Studi di Napoli ``Federico II'',\\
Complesso Universitario Monte Sant'Angelo,\\
Via Cintia, I-80126 Napoli, Italy}
\date{\today}
\maketitle

\begin{abstract}
Following the ideas behind the Feynman approach, a variational wave function 
is proposed for the Fr\"ohlich model. 
It is shown that it provides, for 
any value of the electron-phonon coupling constant, an  
estimate of the polaron ground state energy better than the Feynman 
method based on path integrals. The mean number of phonons, 
the average electronic kinetic and interaction 
energies, the ground state spectral weight and the 
electron-lattice correlation function are calculated and 
successfully compared with the best available results. 

\end {abstract} 

\newpage
\section {Introduction}

In recent years a large amount of experimental results 
has pointed out that the electron-phonon (e-ph) 
interaction plays a significant role in determining the electronic 
and magnetic properties of new materials as 
the high $T_c$ superconductors and the 
colossal magneto-resistance manganites.\cite{1} The experimental data 
have given rise to a renewed interest in models of the e-ph coupled 
system. In this paper we investigate  
the polaronic features of the Fr\"ohlich model
within a variational approach.\cite{2} Here the picture is 
the following. When an electron in the conduction band of a polar 
crystal moves through the crystal, its Coulomb field produces in  its 
neighborhood an ionic polarization that will influence 
the electron motion. Then the particle must carry this polarization 
with it during its motion through the crystal. The quasi-particle 
formed by the electron and the induced polarization charge is called 
polaron. Within the Fr\"ohlich model: 1) the optical modes have the 
same frequency; 2) the dielectric is treated as a continuum medium; 3) 
in the undistorted lattice the electron moves as a free particle with a 
quadratic dispersion relation (effective band mass approximation). 

The problem of finding the ground state energy of the  Fr\"ohlich Hamiltonian 
attracted the interest of a lot of researchers mainly in the 
period 1950-1955. 
Numerous mathematical techniques have been used to solve this problem:  
from the perturbation theory in the weak coupling regime\cite{3} to the strong 
coupling theory,\cite{4} from the linked cluster theory\cite{5} to 
variational\cite{6} and Monte Carlo approaches.\cite{7,8} 
The weak coupling regime is well described within the  
Lee, Low and Pines (LLP) approach.\cite{9} Here, 
after the dependence of the Hamiltonian on the electron coordinates 
has been eliminated, 
an upper bound for the polaron ground state energy is obtained by using a 
variational 
wave function which is based on the physical assumption that successive 
virtual phonons are emitted independently. In the opposite regime, when the 
e-ph interaction is very strong, a good description of the polaron features 
has been obtained by Landau and Pekar.\cite{10} 
Their theory, based on a variational 
calculation, stems from the idea that, for 
very large values of the e-ph coupling constant, the electron can follow 
adiabatically the quantum zero-point fluctuations of the polarization field. 
In their  
first papers the electron is localized with a Gaussian wave function. 
Next, the method has been 
improved by Hohler\cite{11} 
by constructing an eigenstate of the total wave number 
by superposing Landau-Pekar states localized at different points of the 
lattice. In any case 
the validity of LLP and Hohler approaches is restricted, respectively, 
to weak and strong e-ph coupling regimes.     

An excellent approximation, that is accurate at 
all couplings, has been introduced by Feynman.\cite{12} 
His approach provides a variational 
estimate of the electron self-energy based on the path integrals and the 
Feynman-Jensen inequality. After the phonon variables have been 
eliminated exactly, 
Feynman introduces a model Hamiltonian which describes approximatively 
the interaction of the electron with the lattice. This Hamiltonian is that 
of an electron coupled to another particle with a harmonic oscillator 
coupling. 
The trial action for the system is obtained by eliminating the coordinates 
of the fictitious particle simulating the phonon degrees of freedom. The mass 
$M$ of the fictitious particle and the spring constant are the two variational 
parameters within the Feynman approach. The Monte Carlo study\cite{7,8} of the 
Fr\"ohlich model has demonstrated the remarkable accuracy of the Feynman 
method to the electron self-energy. 

In this paper we use a variational 
technique, within an Hamiltonian approach, 
to investigate the polaronic features of the Fr\"ohlich model. 
It is based on linear superposition of two translationally invariant wave 
functions that provide a very good description of the weak and 
strong e-ph coupling regimes. 
These wave functions are built assuming as starting points the LLP\cite{9}
and Hohler\cite{11} variational approaches. 
First, we improve these methods obtaining 
a better upper bound for the polaron ground state energy in the two 
asymptotic regimes of weak and strong e-ph interaction. Then, we use a linear 
superposition of these two wave functions. The comparison of our results 
with the Feynman\cite{12} and Monte Carlo data\cite{7} 
shows that the proposed method provides
an excellent description of the polaron ground state energy for any value 
of the e-ph coupling. Within our variational approach, the estimate of the 
electron self-energy turns out systematically lower than one of the Feynman 
method. In particular, unlike the Feynman approach, the ground state energy 
shows the exact dependence on the e-ph coupling constant in the strong 
coupling regime. Next, we calculate the mean number of phonons present in 
the virtual phonon cloud surrounding the electron, the average electronic 
kinetic and interaction energies, 
the ground state spectral weight and the induced ionic polarization charge 
density. These quantities are successfully compared with Monte Carlo\cite{8} 
and Feynman results.\cite{13}

The proposed method has the advantage to exhibit, first 
to the author's knowledge, a wave function that gives the correct behavior 
in both weak and strong coupling limits and provides an interpolation 
between them with results at least accurate as those of the Feynman approach.
\cite{12}

\section {The model}

The Fr\"ohlich model\cite{2} is described by the Hamiltonian: 

\begin {equation}
H=H_{el}+H_{ph}+H_{e-ph}=\frac{p^2}{2m}+ 
\sum_{\vec{q}}\hbar\omega_{0}a^{\dagger}_{\vec{q}}a_{\vec{q}}+ 
\sum_{\vec{q}}(M_qe^{i\vec{q}\cdot\vec{r}}a_{\vec{q}}+h.c.).
\label{1r} 
\end {equation} 
	
In Eq.(\ref{1r}) $m$ is the band mass of the electron, $\hbar\omega_{0}$ 
is the longitudinal optical phonon energy, $\vec{r}$ and $\vec{p}$ are the 
position and momentum operators of the electron, $a^{\dagger}_{\vec{q}}$ 
represents the creation operator for phonons with wave number $\vec{q}$ and 
$M_q$ indicates the e-ph matrix element.  
In the Fr\"ohlich model,\cite{2} $M_q$ assumes the form:  
\begin {equation} 
M_{q}=i\hbar\omega_{0}\frac{R^{1/2}_{p}}{q}\sqrt{\frac{4\pi\alpha}{V}}~,
\label{2r} 
\end {equation} 
where $\alpha$, dimensionless quantity, is the e-ph coupling constant, 
$R_{p}=\sqrt{\frac{\hbar}{2m\omega_{0}}}$ and $V$ is the 
volume of the system. 

\section {The strong coupling regime}

\subsection{The adiabatic approximation}

When the value of $\alpha$ is very large ($\alpha\gg1$) the electron can 
follow adiabatically the lattice polarization changes and it becomes 
self-trapped in the induced polarization field. The idea of 
Landau and Pekar,\cite{10} 
in the first works on polarons, is to use, as trial wave function for the e-ph 
coupled system, a product of normalized variational wave functions 
$|\varphi\rangle$ and $|f\rangle$ depending, respectively, 
on the electron and phonon coordinates:
\begin{equation}
|\psi\rangle=|\varphi\rangle|f\rangle.
\label{12r}
\end{equation}   

The expectation value of the Hamiltonian (\ref{1r}) on the state 
(\ref{12r}) gives: 
\begin{equation}
\langle\psi|H|\psi\rangle=\langle\varphi|\frac{p^2}{2m}|\varphi\rangle+
\langle f|\sum_{\vec{q}}\left[
\hbar\omega_0a^{\dagger}_{\vec{q}}a_{\vec{q}}+\rho_{\vec{q}}a_{\vec{q}}
+\rho^*_{\vec{q}}a^{\dagger}_{\vec{q}}
\right]|f\rangle 
\label{13r}
\end{equation} 
with 
\begin{equation}
\rho_{\vec{q}}=M_{q} \langle\varphi |e^{i\vec{q}\cdot \vec{r}}|\varphi
\rangle.
\label{14r}
\end{equation}

The variational problem with respect to $|f>$ leads to the following 
lowest energy phonon state: 
\begin{equation}
|f>=e^{\sum_{\vec{q}}\left[ \frac{\rho_{\vec{q}}}{\hbar\omega_0}a_{\vec{q}}
-h.c.\right]}|0>.
\label{15r}
\end{equation} 
The minimization of the corresponding energy with respect to $|\varphi\rangle$ 
leads to a non-linear integro differential equation that has been solved 
numerically by Miyake.\cite{17} 
The result for the polaron ground state energy in the 
strong coupling limit is: 
\begin {equation} 
E=-0.108513\alpha^2\hbar\omega_0. 
\label{17r} 
\end {equation} 

The Landau-Pekar\cite{10} Gaussian ansatz for $|\varphi\rangle$: 
\begin {equation}  
|\varphi_{lp}\rangle=e^{-\left(\frac{m\omega}{\hbar}\right)^2\frac{r^2}{2}} 
\left(\frac{m\omega}{\hbar\pi}\right)^{3/4},
\label{18r} 
\end {equation} 
after the minimization of the expectation value of the Hamiltonian (\ref{1r}) 
on this state, 
with respect to the variational parameter $\omega$, provides an estimate 
of the ground state energy: 
\begin {equation} 
E=-\frac{\alpha^2}{3\pi}\hbar\omega_0  \simeq -0.106103\alpha^2\hbar\omega_0 
\label{19r} 
\end {equation} 
that is very close to the exact result (\ref{17r}).
The best value for $\omega$ turns out: 
\begin {equation} 
\omega=\frac{4\alpha^2}{9\pi}\omega_0. 
\label{20r} 
\end {equation}
 
An excellent approximation for the true energy (\ref{17r}) is obtained by 
using a trial wave function 
similar to that one introduced by Pekar:\cite{10} 
\begin {equation}
|\varphi_{p}\rangle=N e^{-\gamma r} \left[ 1+ b \left(2 \gamma r\right)+c 
\left(2 \gamma r \right)^2 \right] 
\label{21r} 
\end {equation}
with $N$ normalization constant and $b$, $c$ and $\gamma$ variational 
parameters. The minimization of $\langle\varphi_p|H|\varphi_p\rangle$ leads 
to: 
\begin {equation} 
E=-0.108507\alpha^2\hbar\omega_0. 
\label{22r} 
\end {equation}
This upper bound for the energy differs from the exact value less than 
$0,01\%$. 

\subsection{Path integral method versus Hamiltonian approach}

At this point we recall the result of the Feynman\cite{12} 
variational calculation 
when the approximating action is represented by a fixed harmonic binding 
potential: 
\begin {equation} 
E=\left[-\frac{\alpha^2}{3\pi}-3\log{2}\right]\hbar \omega_0, 
 ~~~~ \alpha\rightarrow\infty.
\label{23r} 
\end {equation}

It is given by the sum of two terms. 
The first one corresponds to use  
a Gaussian wave function in the Landau and Pekar's method (Eq.(\ref{18r})). 
The last one does not depend on the e-ph coupling constant $\alpha$. 
The origin 
of this contribution in Feynman's expansion of the polaron energy has been 
discussed by Allcock\cite{18} 
by using the perturbation theory in the strong coupling 
limit. Our first aim is to put this result on variational basis. 
This will allow us to characterize the terms that one has to introduce in 
the trial wave function to improve the Landau and Pekar's ansatz. To this aim, 
starting from Eq.(\ref{12r}) and Eq. (\ref{15r}), we apply the following 
unitary transformation: 
\begin {equation} 
H_1=e^{S_{1}}He^{-S_{1}} 
\label{24r} 
\end {equation}
with
\begin {equation} 
S_{1}=-\sum_{\vec{q}}\left[\frac{\alpha_{\vec{q}}}{\hbar\omega_0}a_{\vec{q}}-
h.c.\right].
\label{25r} 
\end {equation}
The transformed Hamiltonian assumes the form: 
\begin {equation} 
H_1=H_{0}+H_{I} 
\label{26r} 
\end {equation}
with
\begin {equation} 
H_{0}= \frac{p^2}{2m}+ 
\sum_{\vec{q}}\hbar\omega_{0}a^{\dagger}_{\vec{q}}a_{\vec{q}}-
\sum_{\vec{q}}\left[\left(M_qe^{i\vec{q}\cdot\vec{r}}-\alpha_{\vec{q}}\right)
\frac{\alpha^{*}_{\vec{q}}}{\hbar\omega_0}
+h.c.\right]-\sum_{\vec{q}}\frac{|\alpha_{\vec{q}}|^2}
{\hbar\omega_0}
\label{rbis} 
\end {equation}
and 
\begin {equation}
H_{I}=\sum_{\vec{q}}\left[\left(M_qe^{i\vec{q}\cdot\vec{r}}-\alpha_{\vec{q}}
\right)a_{\vec{q}}+h.c.\right].
\label{27r} 
\end {equation}
One recognizes immediately that the Landau-Pekar approach corresponds to use  
as trial wave function for $H_1$: 
\begin {equation}
|\psi\rangle^{(0)}=|0\rangle|\varphi_{lp}\rangle
\label{28r} 
\end {equation}
with $|\varphi_{lp}\rangle$ given by Eq.(\ref{18r}) and 
$\alpha_q=M_q\langle\varphi_{lp} 
|e^{i\vec{q}\cdot \vec{r}}|\varphi_{lp}\rangle$. 
In other words, in this approach, one approximates the lowest energy state 
of $H_{0}$ with a Gaussian wave function containing the variational parameter 
$\omega$, that represents the characteristic oscillation of the electron 
in the induced lattice polarization. The next order term is 
obtained assuming $H_{I}$ as 
perturbation and approximating the eigenstates of $H_{0}$ with those of an 
harmonic oscillator. At the first order of the perturbation theory the wave 
function is:
\begin{equation}
|\psi\rangle^{(1)}=|\psi\rangle^{(0)}-\int_{0}^{1} 
 t^{\left[\frac{\omega_0}{\omega}-1\right]}  
\sum_{\vec{q}}h^{*}_{\vec{q}}(\vec{r},t) a^{\dagger}_{\vec{q}}|0\rangle|
\varphi_{lp}\rangle dt
\label{29r} 
\end {equation}
where
\begin{equation}
h_{\vec{q}}(\vec{r},t)=\frac{M_q}{\hbar\omega} 
e^{i\vec{q}\cdot\vec{r} t} e^{\frac{q^2}{2}\frac{\hbar}{2m\omega}\left(
t^2-1\right)}-\frac{\alpha_{\vec{q}}}{\hbar\omega}~.
\label{30r} 
\end {equation}
This expression has been got using the generating function of the 
Hermite polynomials. Finally we note that $|\psi\rangle^{(1)}$ can be obtained 
from 
\begin{equation}
|\psi\rangle=e^{-S_{2}}|\varphi_{lp}\rangle|0\rangle
\label{31r} 
\end {equation}
with
\begin{equation}
S_{2}=-\int_{0}^{1} t^{\left[\frac{\omega_0}{\omega}-1\right]}
\sum_{\vec{q}}\left[h_{\vec{q}}(\vec{r},t)a_{\vec{q}}-h.c.\right]
dt~,
\label{32r} 
\end {equation}
by expanding the exponential $e^{-S_{2}}$ and truncating the expansion 
at the first order. Taking into account also the unitary transformation 
in Eq.(\ref{24r}), the previous considerations lead us to assume as trial 
wave function for the Fr\"ohlich Hamiltonian in the strong coupling limit: 
\begin{equation}
|\psi_{F}\rangle=e^{-S_{F}}|\varphi_{lp}\rangle|0\rangle,
\label{33r} 
\end {equation}
with
\begin{equation}
S_{F}=-\int_{0}^{1}t^{\left[\frac{\omega_0}{\omega}-1\right]}
\sum_{\vec{q}}\left[
\frac{M_q}{\hbar\omega} 
e^{i\vec{q}\cdot\vec{r} t} e^{\frac{q^2}{2}\frac{\hbar}{2m\omega}\left(
t^2-1\right)}
a_{\vec{q}}-h.c.\right]
dt~.
\label{34r} 
\end {equation}
We have indicated this coherent state with ''F'' since it is easy to show 
that the expectation value of the  Fr\"ohlich Hamiltonian on the 
state (\ref{34r}) gives: 
\begin{equation}
E=\frac{3}{4}\hbar\omega-\alpha\hbar\omega_0\sqrt{\frac{\omega_0}{\omega}}
\frac{\Gamma(\frac{\omega_0}{\omega})}
{\Gamma(\frac{\omega_0}{\omega}+\frac{1}{2})}~,
\label{35r} 
\end {equation}
i.e. the Feynman result when the approximating action is represented 
by a fixed harmonic binding potential.\cite{12} 
In Eq.(\ref{35r}) $\Gamma(x)$ is 
the Gamma function. In particular the minimization of $E$ with respect to 
the variational parameter $\omega$ and the asymptotic expansion for 
$\alpha\rightarrow\infty$ restore the Eq.(\ref{23r}). Then, the order 
beyond  the Landau and Pekar's theory is due to the lattice fluctuations 
and to the consequent change in the electron wave function. 

\subsection{Improvements of the Feynman result}

The next step is to try to improve the Feynman result. To this aim, we note 
that is possible to obtain an excellent approximation of the polaron 
ground state energy in Eq.(\ref{35r}) 
substituting in Eq.(\ref{34r}) $S_{F}$ with:
\begin{equation}
S=-\sum_{\vec{q}}
\left[\left(
v_{\vec{q}}e^{i\vec{q}\cdot \vec{r}\eta}+
u_{\vec{q}}e^{i\vec{q}\cdot \vec{r}}
\right)
a_{\vec{q}}-h.c.\right]
\label{36r} 
\end {equation}
where
\begin{equation}
v_{\vec{q}}=\frac{M_q}{\hbar\omega}
\int_{0}^{a}t^{\left[\frac{\omega_0}{\omega}-1\right]}
e^{\frac{q^2}{2}\frac{\hbar}{2m\omega}\left(
t^2-1\right)}
\label{37r} 
\end {equation}
and 
\begin{equation}
u_{\vec{q}}=\frac{M_q}{\hbar\omega}
\int_{a}^{1}t^{\left[\frac{\omega_0}{\omega}-1\right]}
e^{\frac{q^2}{2}\frac{\hbar}{2m\omega}\left(
t^2-1\right)}~.
\label{38r} 
\end {equation}
Here $a$ and $\eta$ are two variational parameters. In other words, we obtain 
the main contribution to the Feynman estimate of the electron self-energy 
approximating $S_{F}$ as sum of two terms: the first one stems from the 
observation that the electron moves very fast in the induced potential 
well; the second one takes into account the lattice fluctuations and the 
possibility that they can follow instantaneously the electron motion. 
In order to improve the Feynman result, the Pekar's approach 
(Eq.(\ref{21r})) and the previous analysis suggest us to try the 
following ansatz:
\begin{equation}
|\psi\rangle=e^{
-\sum_{\vec{q}}
\left[\left(
s_{\vec{q}}e^{i\vec{q}\cdot \vec{r}}+
l_{\vec{q}}e^{i\vec{q}\cdot \vec{r}\eta}
\right)
a_{\vec{q}}-h.c.\right]
}|0\rangle|\varphi_{p}\rangle
\label{39r}
\end{equation}
with $|\varphi_{p}\rangle$ given by Eq.(\ref{21r}), $\eta$ variational 
parameter and $l_{\vec{q}}$ and $s_{\vec{q}}$ functions to be determined 
by minimizing the expectation value of the  Fr\"ohlich Hamiltonian 
on this state. This last quantity turns out: 
\begin{eqnarray}
\langle\psi|H|\psi\rangle= 
&&\langle\varphi_{p}|\frac{p^2}{2m}|\varphi_{p}\rangle+
\sum_{\vec{q}}\left[\hbar\omega_0\left(|l_{\vec{q}}|^2+|s_{\vec{q}}|^2
\right)+\frac{\hbar^2 q^2} {2m}\left(\eta^2|l_{\vec{q}}|^2+|s_{\vec{q}}|^2
\right)\right] \nonumber \\
&&+\sum_{\vec{q}}\left[
\left(\hbar\omega_0+\frac{\hbar^2 q^2} {2m}\eta\right)
\left(r_{\vec{q}}s_{\vec{q}}l^{*}_{\vec{q}}+h.c.\right)
-\left(M_qs^{*}_{\vec{q}}+M_qr_{\vec{q}}l^{*}_{\vec{q}}+h.c.\right)
\right]
\label{40r}
\end{eqnarray}
with 
\begin{equation}
r_{\vec{q}}=
\langle\varphi_{p}|e^{i\vec{q}\cdot\vec{r}\left(1-\eta\right)}|\varphi_{p}
\rangle.
\label{41r}
\end{equation}

Making $\langle\psi|H|\psi\rangle$ stationary with respect to arbitrary 
variations of the functions $l_{\vec{q}}$ and $s_{\vec{q}}$, we obtain 
two, easily solvable, algebraic equations. The minimization and the 
asymptotic expansion of the ground state energy, 
for $\alpha\rightarrow\infty$, provide: 
\begin{equation}
E=\left[-0.108507\alpha^2-1.89\right]\hbar\omega_0.
\label{42r}
\end{equation}

The electron self-energy shows the exact dependence on $\alpha^2$ in the 
strong coupling regime together with a good estimate of the e-ph coupling 
constant independent contribution due to the lattice fluctuations. This allows 
to obtain, for $\alpha\ge 8.7$,  
an upper bound for the polaron ground state energy better than the Feynman 
approach when the approximating action is represented by a fixed harmonic 
binding potential (Eq.(\ref{35r})). On the other hand, both these methods 
give the same result for $\alpha\le 6$, i.e. $E=-\alpha\hbar\omega_0$. 
However, both the methods show a discontinuity in the transition from 
the weak to strong coupling regime. 

To overcome this difficulty one has to take into account 
the translational invariance. We construct an eigenstate of the total wave 
number by taking a superposition of the localized states (\ref{39r}) centered 
on any point of the lattice in the same manner in which one constructs 
a Bloch wave function from a linear combination of atomic orbitals: 
\begin{equation}
|\psi_{(sc)}\rangle=\int \psi(\vec{r}-\vec{R}) d^3R~. 
\label{43r}
\end{equation}
The minimization, with respect to the variational parameters, 
of the expectation 
value of the Fr\"ohlich Hamiltonian on this state, that accounts for the 
translationally symmetry, and the asymptotic expansion for 
$\alpha\rightarrow\infty$ provide: 
\begin{equation}
E=\left[-0.108507\alpha^2-2.67\right]\hbar\omega_0.
\label{44r}
\end{equation}
This upper bound is lower than the variational Feynman estimate which for 
large values of $\alpha$ assumes the form:\cite{12} 
\begin {equation} 
E=\left[-\frac{\alpha^2}{3\pi}-3\log{2}-\frac{3}{4}\right]\hbar \omega_0.
\label{45r} 
\end {equation}

\section {The weak coupling regime}

When the value of $\alpha$ is very small the lattice follows adiabatically 
the electron. 
A good physical description of the polaron features in this regime is 
provided by the 
LLP approach.\cite{9} 
The starting point is the observation that the total momentum 
operator: 
\begin {equation} 
\vec{P}_t=\vec{p}+\sum_{\vec{q}}\hbar\vec{q}a^{\dagger}_{\vec{q}}a_{\vec{q}} 
\label{3r} 
\end {equation} 
is a motion constant, i.e. it commutes with the Hamiltonian. The conservation 
law of the 
total momentum is taken into account through the unitary transformation: 
\begin {equation} 
U=e^{i\left(\vec{Q}-\sum_{\vec{q}}\vec{q}a^{\dagger}_{\vec{q}}a_{\vec{q}}
\right)\cdot\vec{r}}~,
\label{4r} 
\end {equation} 
where $\hbar\vec{Q}$ is the eigenvalue of $\vec{P}_t$. In this paper we are 
interested in the ground state properties of the e-ph coupled system, 
so that we will restrict 
ourselves to the case $\vec{Q}=0$. The transformed Hamiltonian does not 
contain the electron variables and it is given by: 
\begin {equation} 
H_1=U^{-1}HU=\sum_{\vec{q}}\left(\hbar\omega_{0}+\frac{\hbar^2q^2}{2m}
\right)a^{\dagger}_{\vec{q}}a_{\vec{q}}+
\sum_{\vec{q}}(M_qa_{\vec{q}}+h.c.)+\frac{\hbar^2}{2m}\sum_{\vec{q}_1,
\vec{q}_2}\vec{q}_1\cdot\vec{q}_2
a^{\dagger}_{\vec{q}_1}a^{\dagger}_{\vec{q}_2}a_{\vec{q}_2}a_{\vec{q}_1}.
\label{5r} 
\end {equation} 
The LLP wave function is:
\begin {equation} 
|\psi\rangle=e^{\sum_{\vec{q}}\left(f_{\vec{q}}a_{\vec{q}}-h.c.\right)}
|0\rangle,
\label{6r} 
\end {equation} 
where $|0\rangle$ is the phonon vacuum state and $f_{\vec{q}}=M_q/\left(
\hbar\omega_0+\frac{\hbar^2 q^2}{2m}\right)$. In other words $|\psi\rangle$ 
is the lowest energy state of the first two terms of the transformed 
Hamiltonian $H_1$. The use of this wave function is based on the physical 
assumption that, when the e-ph interaction is weak, 
there is not correlation among the emission of successive 
virtual phonons by the electron. This 
assumption restricts the validity of this approach to the regime characterized 
by small values of $\alpha$. The ground state energy turns out $E=-\alpha\hbar
\omega_0$. In other words, this method puts the 
results of the perturbation theory on variational basis. 

To improve the LLP approximation,\cite{9} 
one has to introduce in the trial wave function a better description 
of the recoil effect of the electron, effect present only on average 
in LLP approach. This can be done using the following ansatz: 
\begin {equation} 
|\psi_{(wc)}\rangle=e^{\sum_{\vec{q}}\left(g_{\vec{q}}a_{\vec{q}}-h.c.\right)}
\left[|0\rangle+\sum_{\vec{q}_1,\vec{q}_2}d_{\vec{q}_1,\vec{q}_2}
a^{\dagger}_{\vec{q}_1}a^{\dagger}_{\vec{q}_2}|0\rangle\right],
\label{7r} 
\end {equation}  
that takes into account the correlation between the virtual emission  
of pairs of phonons.\cite{14} In this paper we will choose: 
\begin {equation} 
g_{\vec{q}}=\frac{M_q}{\left(
\hbar\omega_0+\frac{\hbar^2 q^2}{2m}\epsilon^2\right)}
\label{8r} 
\end {equation}  
and 
\begin {equation} 
d_{\vec{q}_1,\vec{q}_2}=\gamma \hbar \omega_0 \frac{\hbar^2}{2m} \vec{q}_1
\cdot \vec{q}_2  
\frac{M_{{q}_1}}{\left(
\hbar\omega_0+\frac{\hbar^2 {{q}^2_1}}{2m}\delta^2\right)}
\frac{M_{{q}_2}}{\left(
\hbar\omega_0+\frac{\hbar^2 {{q}^2_2}}{2m}\delta^2\right)}~.
\label{9r} 
\end {equation}  
Here $\gamma$, $\delta$ and $\epsilon$ are three variational parameters that 
have to be determined by minimizing the expectation value of the 
Hamiltonian (\ref{1r}) on the state (\ref{7r}). This procedure provides as 
upper 
bound for the polaron ground state energy at small values of $\alpha$: 
\begin {equation} 
E=-\alpha\hbar\omega_0-0.0123\alpha^2\hbar\omega_0,~~ \alpha\rightarrow 0~,
\label{10r} 
\end {equation} 
i.e. the same result, at this order, of the 
Feynman approach.\cite{12} 
We stress that, at the $\alpha^2$ order, 
the result for the electron self-energy is: 
\begin {equation} 
E=-\alpha\hbar\omega_0-0.0159\alpha^2\hbar\omega_0
\label{11r} 
\end {equation} 
as found by Hohler and Mullensiefen,\cite{15} Larsen\cite{14} and 
Roseler.\cite{16}

\section {All couplings}

A careful inspection of the wave function (\ref{43r}) shows that is able to 
interpolate between strong and weak coupling regimes. On the other hand, for 
small values of $\alpha$ a better description of the polaron ground state 
features is provided by the wave function (\ref{7r}). Moreover, in the 
weak and intermediate e-ph coupling, $\alpha\le 7$, these two solutions 
are not orthogonal and have non zero off diagonal matrix 
elements. This suggests that the lowest state of the system is made of a 
mixture of the two wave functions that give an accurate description 
of weak and strong e-ph coupling regimes.   
Then the idea is to use a variational method to determine the 
ground state energy of the Hamiltonian (\ref{1r}) 
by considering as trial state 
a linear superposition of the two previously discussed wave functions:
\begin{equation}
|\psi\rangle=\frac{A 
|\overline{\psi}_{(wc)}\rangle+
B|\overline{\psi}_{(sc)}\rangle}
{\sqrt{A^2+B^2+2 A B S}},
\label{46r}
\end{equation} 
where 
\begin{eqnarray}
&&|\overline{\psi}_{(wc)}\rangle=
\frac{|\psi_{(wc)}\rangle}
{\sqrt{\langle\psi_{(wc)}|\psi_{(wc)}\rangle}}, 
~~|\overline{\psi}_{(sc)}\rangle=
\frac{|\psi_{(sc)}\rangle}
{\sqrt{\langle\psi_{(sc)}|\psi_{(sc)}\rangle}},
\label{47r}
\end{eqnarray}
and $S$ is the overlap factor:  
\begin{equation}
S=\frac{\langle\overline{\psi}_{(wc)}|\overline{\psi}_{(sc)}\rangle+h.c.}
{2}~.
\label{48r}
\end{equation}
In Eq.(\ref{46r}) $A$ and $B$ 
are two additional variational parameters 
that provide the relative weight of the two 
solutions in the ground state of the system. 
In this paper we perform the minimization procedure in two steps. First,  
the expectation values of  the Fr\"ohlich Hamiltonian on   
the two trial wave functions in Eq.(\ref{7r}) and Eq.(\ref{43r}) are 
minimized and the variational parameters are determined. 
Then, the minimization 
procedure discussed in the present section is carried out. 
This way to proceed simplifies 
significantly the computational effort and makes all described 
calculations accessible on a personal computer. An approach, similar 
to that one described in this section, has been successfully used for the 
Holstein model.\cite{us} 

The procedure of minimization of the quantity 
$\langle\psi|H|\psi\rangle$ 
with respect to $A$ and $B$ 
gives for the polaron ground state energy 
\begin{equation}
E=\frac{E_m-SE_c-\sqrt{\left(E_m-SE_c\right)^2-
\left(1-S^2\right)
\left(E_{(wc)}E_{(sc)}-E^2_c\right)}
}{1-S^2}
\label{49r}
\end{equation} 
and for the ratio of the two parameters $A$ and $B$
\begin{equation}
\frac{A}{B}=
\frac{E_c-ES}
{E-E_{(wc)}}~.
\label{50r}
\end{equation}
Here $E_{(wc)}=\langle\overline{\psi}_{(wc)}|H|\overline{\psi}_{(wc)}\rangle$, 
$E_{(sc)}=\langle\overline{\psi}_{(sc)}|H|\overline{\psi}_{(sc)}\rangle$,
$E_m=\left(E_{(wc)}+E_{(sc)}\right)/2$ and  
$E_c=\left(\langle\overline{\psi}_{(wc)}|H|
\overline{\psi}_{(sc)}\rangle+h.c.\right)/2$. 

\section {Numerical results}
In Fig.1 we plot the polaron ground state energy, 
obtained within our approach, as a function of the e-ph coupling constant 
$\alpha$. The data are compared with the results of the variational treatments 
due to Lee, Low and Pines,\cite{9} Pekar,\cite{10} 
Feynman\cite{12} and with the energies calculated 
within a diagrammatic Quantum Monte-Carlo method.\cite{7} 
As it is clear from the plots,  
our variational proposal recovers the asymptotic result of the 
Feynman approach in the weak coupling regime, improves the Feynman's data 
particularly in the opposite regime, characterized by values of the e-ph 
coupling constant $\alpha\gg 1$, and it is in very good agreement with the 
best available results in literature, obtained with the Quantum Monte 
Carlo calculation.\cite{7} 
This agreement indicates that the true ground state 
wave function is very close to a superposition of the above introduced 
functions, that provide a very good description of the two asymptotic 
regimes.  Within our approach we have also calculated the mean number 
of phonons present in the virtual phonon cloud surrounding the electron, $N$, 
the average electronic kinetic and interaction energies, $K$ and $I$. These 
quantities are reported, respectively, in Fig.2, Fig.3 and Fig.4 
where are compared with the same properties 
obtained in the Feynman's variational treatment based on path integrals.
\cite{13} 
As for the ground state energy, our variational ansatz is able to recover 
all the expected behaviors. For small values of $\alpha$, $N\rightarrow 
\alpha/2$, $K\rightarrow \alpha\hbar\omega_0/2$,  
$I\rightarrow -2\alpha\hbar\omega_0$ as predicted by the LLP approach\cite{9} 
and 
the weak coupling perturbation theory.\cite{15} 
In the opposite regime the electron 
follows adiabatically the lattice polarization changes. 
The values $N=
\frac{2\alpha^2}{3\pi}$,  $K=\frac{\alpha^2}{3\pi}\hbar\omega_0$, 
$I=-\frac{4\alpha^2}{3\pi}\hbar\omega_0$ obtained within the Landau and 
Pekar's variational treatment (see Eq.(\ref{18r})), 
based on the electron self-trapping with a Gaussian wave function,  
represent very 
accurate estimates of these quantities when they are 
calculated within the Feynman's approach. 
On the other hand, the values $N=
2\cdot 0.108507 \alpha^2$,  $K= 0.108507 \alpha^2 \hbar\omega_0$, 
$I=-4\cdot 0.108507 \alpha^2\hbar\omega_0$ obtained within the Pekar's 
variational treatment (see Eq.(\ref{21r})) represent very 
good approximations for 
the same quantities calculated within our approach. Then the  
variational Feynman's and our methods differ mainly in the 
strong coupling regime as it turns out from the plots in Fig.1, Fig.2, Fig.3 
and Fig.4. We stress that, in this range of values of $\alpha$, our approach 
provides a better estimate of the polaron ground state energy than the 
Feynman's method.\cite{12} 
 
Another physical quantity of interest is $\rho(\vec{r})$, i.e. the average 
ionic polarization charge density induced at a distance $r$ by the electron. 
This quantity is related to the static correlation function between the 
electron position $\vec{r_e}=0$ and the oscillator displacement at $\vec{r}$: 
\begin{equation}
\rho(\vec{r})=-\left(\frac{1}{4\pi e}\right)\langle \psi(\vec{r_e}=0)
|\sum_{\vec{q}}(M_qe^{i\vec{q}\cdot\vec{r}}q^2a_{\vec{q}}+h.c.)|
\psi(\vec{r_e}=0)\rangle.
\label{51r}
\end{equation}
It easy to show, analytically,  
that the exact sum rule for the total induced charge:\cite{19}
\begin{equation}
\int \rho(\vec{r}) d^3r=e\left(\frac{1}{\epsilon_{\infty}}-
\frac{1}{\epsilon_{0}}\right)
\label{52r}
\end{equation}
is satisfied within our variational approach. Figure 5 shows 
$\rho(\vec{r})/\int\rho(\vec{r}) d^3r$ as a function of $r$ for different 
values of the e-ph matrix element $\alpha$ corresponding to weak, intermediate 
and strong coupling regimes. Our data are compared with results obtained 
within the Feynman's method\cite{13} and a path integral Monte Carlo scheme.
\cite{19} If 
the e-ph coupling is weak, the lattice deformation is not able to trap the 
charge carrier. The extension of the polaron is large compared with the 
characteristic length $\sqrt{\frac{\hbar}{2m\omega_0}}$. The situation is 
different in the opposite regime where the lattice deformation is localized 
around the electron. In any case also this correlation function, evaluated 
within our approach, is in agreement with the best data available in 
literature. 

Finally Figure 6 shows the ground state spectral weight: 
\begin{equation}
Z=|\langle\psi|c^{\dagger}_{\vec{k}=0}|0\rangle|^2,
\label{53r}
\end{equation} 
where $|0\rangle$ is the electronic vacuum state containing no phonons and 
$c^{\dagger}_{\vec{k}}$ is the electron creator operator in the momentum 
space.  $Z$ represents the renormalization coefficient of the one-electron 
Green's function and gives the fraction of the bare electron state in the 
polaron trial wave function. This quantity is compared with that one obtained 
in the diagrammatic Quantum Monte Carlo method.\cite{7} 
The result of the weak coupling perturbation theory
is also indicated: $Z=1-\alpha/2$. For small 
values of $\alpha$ the main part of the spectral weight is located at energies 
that correspond approximatively to the bare electronic levels. Increasing the 
e-ph interaction, the spectral weight decreases very fast and becomes 
practically zero in the strong coupling regime. Here the most part of the 
spectral weight is located at excited states. The diagrammatic Quantum 
Monte Carlo study\cite{7} 
of the  Fr\"ohlich polaron has pointed out that there is 
no stable excited states in the energy gap between the ground state energy 
and the continuum. There are, instead, 
several many phonon unstable states at fixed 
energies: $E_{f}-E_{0}\simeq 1,3.5$ and $8.5 \hbar \omega_0$. These results  
seem to be contrary to the data about the optical absorption of 
large polarons,\cite{20} which show, for large values of $\alpha$, 
the presence of a very narrow peak 
corresponding to the electronic transitions from the ground state to the 
first relaxed excited state (RES). The nature of the excited states and the 
optical absorption of polarons in the  Fr\"ohlich model require further study 
which is beyond the scope of this paper. 

In conclusion, in this paper, a variational approach has been developed to 
investigate the features of the Fr\"ohlich model. It has been shown that a 
linear superposition of two wave functions, that describe the two asymptotic 
regimes of weak and strong e-ph coupling, provides an estimate of the polaron 
ground state energy which is in very good agreement with the best available 
results for any value of the e-ph matrix element. All the evaluated 
ground state properties show that the crossover between the two asymptotic 
regimes is very smooth. On the other hand the transfer of spectral weight 
from the polaron ground state to the higher energy bands turns out very fast. 
We stress that, to the best of our knowledge, it is the first time that a 
variational wave function, able to interpolate between the weak and strong 
e-ph coupling regimes, at least carefully as the Feynman method,\cite{12} 
is exhibited for the Fr\"ohlich model. 

\section*{Figure captions} 
\begin {description} 
\item{Fig.1.} 
(a) The polaron ground state energy, $E$, 
is reported as function of 
$\alpha$ in units of $\hbar\omega_0$. 
The data (solid line), obtained within the approach discussed in this paper,  
are compared with the results (diamonds) of the Feynman approach, $E_{F}$, 
and the results (stars) of the 
diagrammatic Quantum Monte-Carlo method, $E_{MC}$, 
kindly provided by A.S. Mishchenko. 
(b) The differences: $E-E_{F}$ (diamonds) and $E-E_{MC}$ (stars) 
are reported as function of $\alpha$.

\item{Fig.2.} 
(a) The mean number of phonons, $N$, 
is plotted as function of $\alpha$. The 
data, obtained within the approach discussed in this paper (solid line),  
are compared with the results of the Feynman approach, $N_{F}$ (diamond), and
the results of the 
diagrammatic Quantum Monte-Carlo method, $N_{MC}$ (stars), 
extracted from Fig.8 of ref.7.
(b) The differences: 
$N_{F}-N$ (diamonds) and $N_{MC}-N$ (stars) 
are reported as function of $\alpha$. The error bars are due to uncertainty 
in the procedure used to extract the numerical values from Fig.8.

\item{Fig.3.}
(a) The average electronic kinetic energy, $K$,  
is plotted as function of $\alpha$  in units of $\hbar\omega_0$. 
The data (solid line), obtained within the approach discussed in this paper,  
are compared with the results (diamonds) of the Feynman approach, $K_{F}$. 
(b) The difference: $K_{F}-K$ (stars)  
is reported as function of $\alpha$.

\item{Fig.4.} 
(a) The average electronic interaction energy, $I$,  
is plotted as function of $\alpha$  in units of $\hbar\omega_0$. 
The data (solid line), obtained within the approach discussed in this paper,  
are compared with the results (diamonds) of the Feynman approach, $I_{F}$. 
(b) The difference: $I_{F}-I$ (stars)  
is reported as function of $\alpha$. 

\item{Fig.5.} 
The average normalized  
ionic polarization charge density, induced at a distance $r$ by the electron, 
is reported for three different values of $\alpha$: (a) $\alpha=1$; 
(b) $\alpha=6$; (c) $\alpha=12$. The data (solid line), 
obtained within the approach discussed in this paper, are compared 
with the results (dashed line) of the Feynman approach  
and the results (dotted line) of the Monte-Carlo method, 
kindly provided by S. Ciuchi. The distance $r$ is measured in units 
of $R_{p}$.

\item{Fig.6.}
The ground state spectral weight, $Z$,  
is plotted as function of $\alpha$. The 
data (solid line), obtained within the approach discussed in this paper,  
are compared with the results (stars) of 
of the diagrammatic Quantum Monte-Carlo method. The result 
of the weak coupling perturbation theory (dashed line) is also indicated. 

\end {description}


\begin{references} 
\bibitem {1} A. Damascelli, Z. Hussain, Z.X. Shen, Rev. Mod. Phys. {\bf 75}, 
473 (2003); S. Lupi, P. Maselli, M. Capizzi, P. Calvani, P. Giura, and P. Roy
Phys. Rev. Lett. {\bf 83}, 4852-4855 (1999);
A.J. Millis, Nature {London}, {\bf 392}, 147 (1998).
\bibitem {2} H. Fr\"ohlich et al., Philos. Mag. {\bf 41}, 221 
(1950); H. Fr\"ohlich, in {\it Polarons and Excitons}, edited by C.G. Kuper 
and G.A. Whitfield (Oliver and Boyd, Edinburg, 1963), pp. 1; 
for a review see: J.T. Devreese, Polarons, in: G.L. Trigg (ed.) 
{\it Encyclopedia of Applied physics}, New York: VCH, vol. 14, pp. 383 (1996) 
and references therein. 
\bibitem {3} G. H\"ohler and A.M. Mullensiefen, 
Z. Physik {\bf 157}, 159 (1959); 
D.M. Larsen, Phys. Rev. {\bf 144}, 697 (1966); 
G.D. Mahan, in {\it Polarons in Ionic Crystals 
and Polar Semiconductors} (North-Holland, Amsterdam, 1972), p. 553; 
M.A. Smondyrev, Teor. Mat. Fiz. {\bf 68}, 29 (1986). 
\bibitem {4} L.D. Landau, Phys. Z. Sowjetunion {\bf 3}, 664 (1933) [
English translation: {\it Collected Papers} (Gordon and Breach New York, 
1965), pp.67-68];  
L.D. Landau and S.I. Pekar, 
J. Exptl. Theor. Phys. {\bf 18}, 419 (1948);
S.I. Pekar, Zh. Eksp. i Theor. Fiz. {\bf 19}, 796 (1949); 
N.N. Bogoliubov and S.V. Tiablikov, Zh. Eksp. i Theor. Fiz. 
{\bf 19}, 256 (1949); 
S.I. Pekar, Research in Electron Theory of Crystals, 
Moscow, Gostekhizdat (1951) [English translation: Research in Electron Theory 
of Crystals, US AEC Report AEC-tr-5575 (1963)];
S.V. Tiablikov, Zh. Eksp. i Theor. Fiz. 
{\bf 21}, 377 (1951); 
S.V. Tiablikov, Abhandl. Sowj. Phys.  
{\bf 4}, 54 (1954);
G.R. Allcock, Advances in Physics {\bf 5}, 412 (1956); 
G. H\"ohler in {\it Field Theory and the Many Body Problem}, ed. E.R. 
Caianiello (Accademic Press), pp.285; E.P. Solodovnikova, A.N. Tavkhelidze, 
O.A. Khrustalev, Teor. Mat. Fiz. {\bf 11}, 217 (1972); V.D. Lakhno and G.N. 
Chuev, Physics-Uspekhi {\bf 38}, 273 (1995).
\bibitem {5} D. Dunn, Can. J. Phys. {\bf 53}, 321 (1975).  
\bibitem {6} T.D. Lee, F. Low, and D. Pines, Phys. Rev. {\bf 90}, 
297 (1953); 
M. Gurari, Phil. Mag. {\bf 44}, 329 (1953);  
E.P. Gross, Phys. Rev. {\bf 100}, 1571 (1955);
G. H\"ohler, Z. Physik {\bf 140}, 192 (1955); 
G. H\"ohler, Z. Physik {\bf 146}, 372 (1956); 
R.P. Feynman, Phys. Rev. {\bf 97}, 660 (1955); 
Y. Osaka, Prog. Theor. Phys. {\bf 22}, 437 (1959); 
T.D. Schultz, Phys. Rev. {\bf 116}, 526 (1959); 
T.D. Schultz, in {\it Polarons and Excitons}, edited by C.G. Kuper 
and  G.A. Whitfield (Oliver and Boyd, Edinburg, 1963), pp. 71; 
D.M. Larsen, Phys. Rev. {\bf 172}, 967 (1968); 
J. Roseler, Phys. Stat. Sol. {\bf 25}, 311 (1968).  
\bibitem {7} A.S. Mishchenko, N.V. Prokof'ev, A. Sakamoto, and B. V. 
Svistunov, Phys. Rev. B {\bf 62}, 6317 (2000); N.V. Prokof'ev and 
B. V. Svistunov, Phys. Rev. Lett. {\bf 81}, 2514 (1998).
\bibitem {8} John T. Titantah, Carlo Pierleoni, and Sergio Ciuchi, 
Phys. Rev. Lett. {\bf 87}, 206406 (2001). 
\bibitem {9} T.D. Lee, F. Low, and D. Pines, Phys. Rev. {\bf 90}, 
297 (1953).
\bibitem {10}  L.D. Landau, Phys. Z. Sowjetunion {\bf 3}, 664 (1933) [
English translation: {\it Collected Papers} (Gordon and Breach New York, 
1965), pp.67-68];  
L.D. Landau and S.I. Pekar, 
J. Exptl. Theor. Phys. {\bf 18}, 419 (1948); 
S.I. Pekar, Research in Electron Theory of Crystals, 
Moscow, Gostekhizdat (1951) [English translation: Research in Electron Theory 
of Crystals, US AEC Report AEC-tr-5575 (1963)]. 
\bibitem {11} 
G. H\"ohler, Z. Physik {\bf 140}, 192 (1955); 
G. H\"ohler, Z. Physik {\bf 146}, 372 (1956).  
\bibitem {12} R.P. Feynman, Phys. Rev. {\bf 97}, 660 (1955).
\bibitem {13} F.M. Peeters, and J.T. Devreese, Phys. Stat. Sol. {\bf 115}, 285 
(1983); 
F.M. Peeters, and J.T. Devreese, Phys. Rev. B {\bf 31}, 4890 (1985).  
\bibitem {17} S.J. Miyake, J. Phys. Soc. Jap. {\bf 38}, 181 (1975); 
S.J. Miyake, J. Phys. Soc. Jap. {\bf 41}, 747 (1976).
\bibitem {18} G.R. Allcock, Advances in Physics {\bf 5}, 412 (1956).
\bibitem {14} D.M. Larsen, Phys. Rev. {\bf 172}, 967 (1968).
\bibitem {15} G. H\"ohler and A.M. Mullensiefen, 
Z. Physik {\bf 157}, 159 (1959).
\bibitem {16}  J. Roseler, Phys. Stat. Sol. {\bf 25}, 311 (1968). 
\bibitem{us} V. Cataudella, G. De Filippis, and G. Iadonisi, Phys. Rev. B 
{\bf 60}, 15163 (1999); 
V. Cataudella, G. De Filippis, and G. Iadonisi, Phys. Rev. B 
{\bf 62}, 1496 (2000).
\bibitem {19} Sergio Ciuchi, 
Jos\'e Lorenzana, and  Carlo Pierleoni, 
Phys. Rev. B {\bf 62}, 4426 (2000).
\bibitem {20} F.M. Peeters, and J.T. Devreese, Phys. Rev. B {\bf 28}, 6051 
(1983).
\end {references}
\end {document}
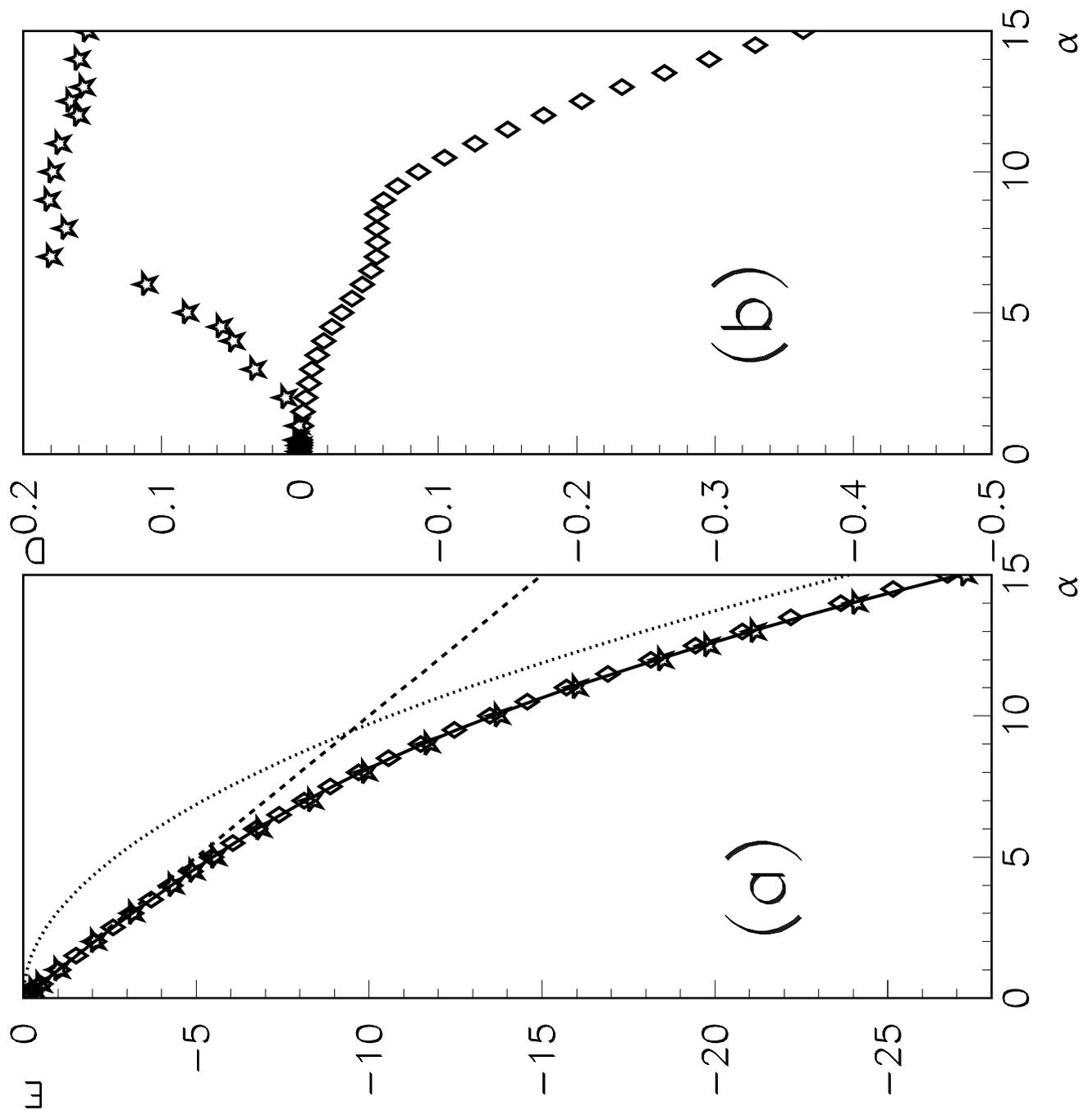
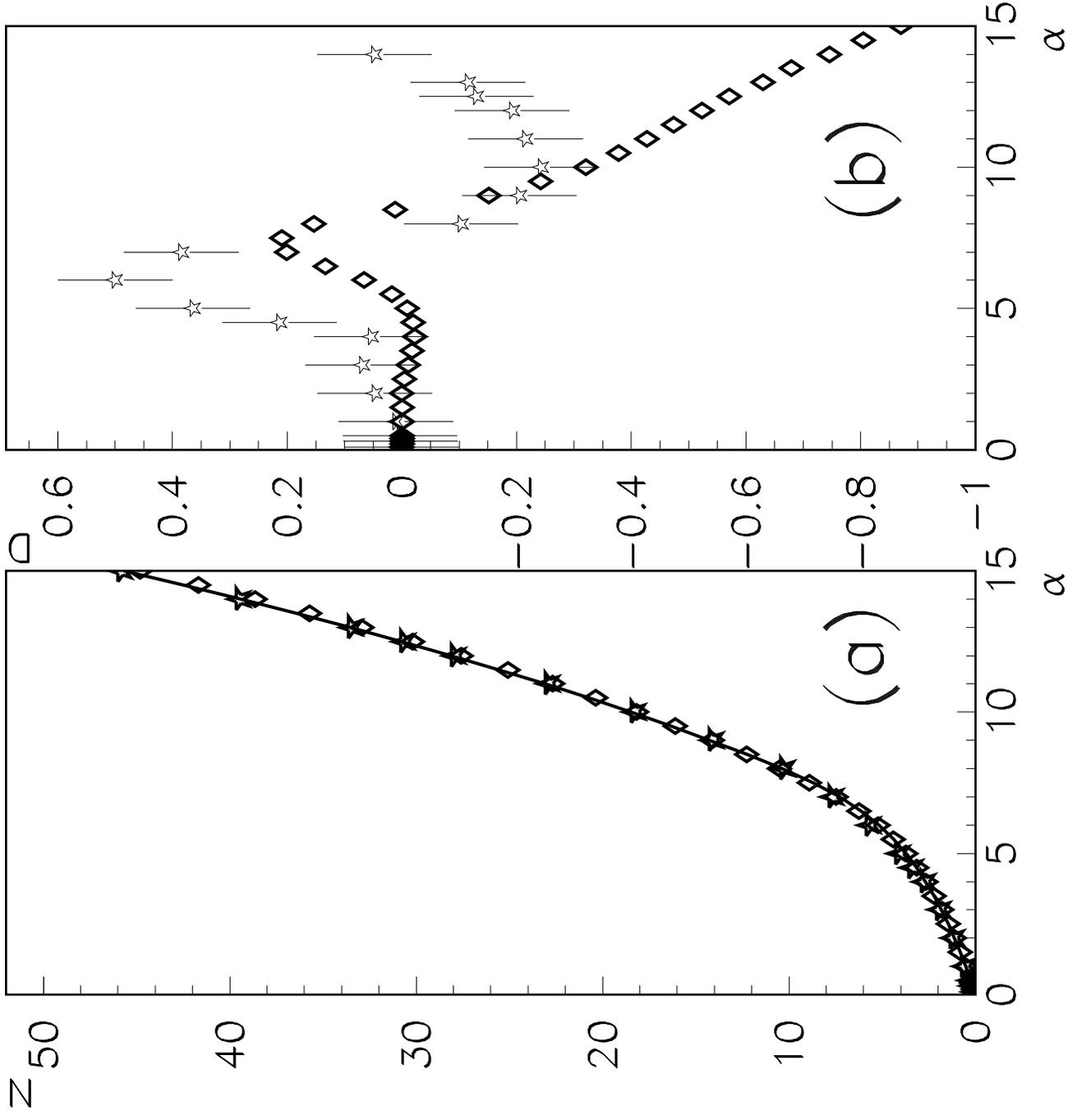
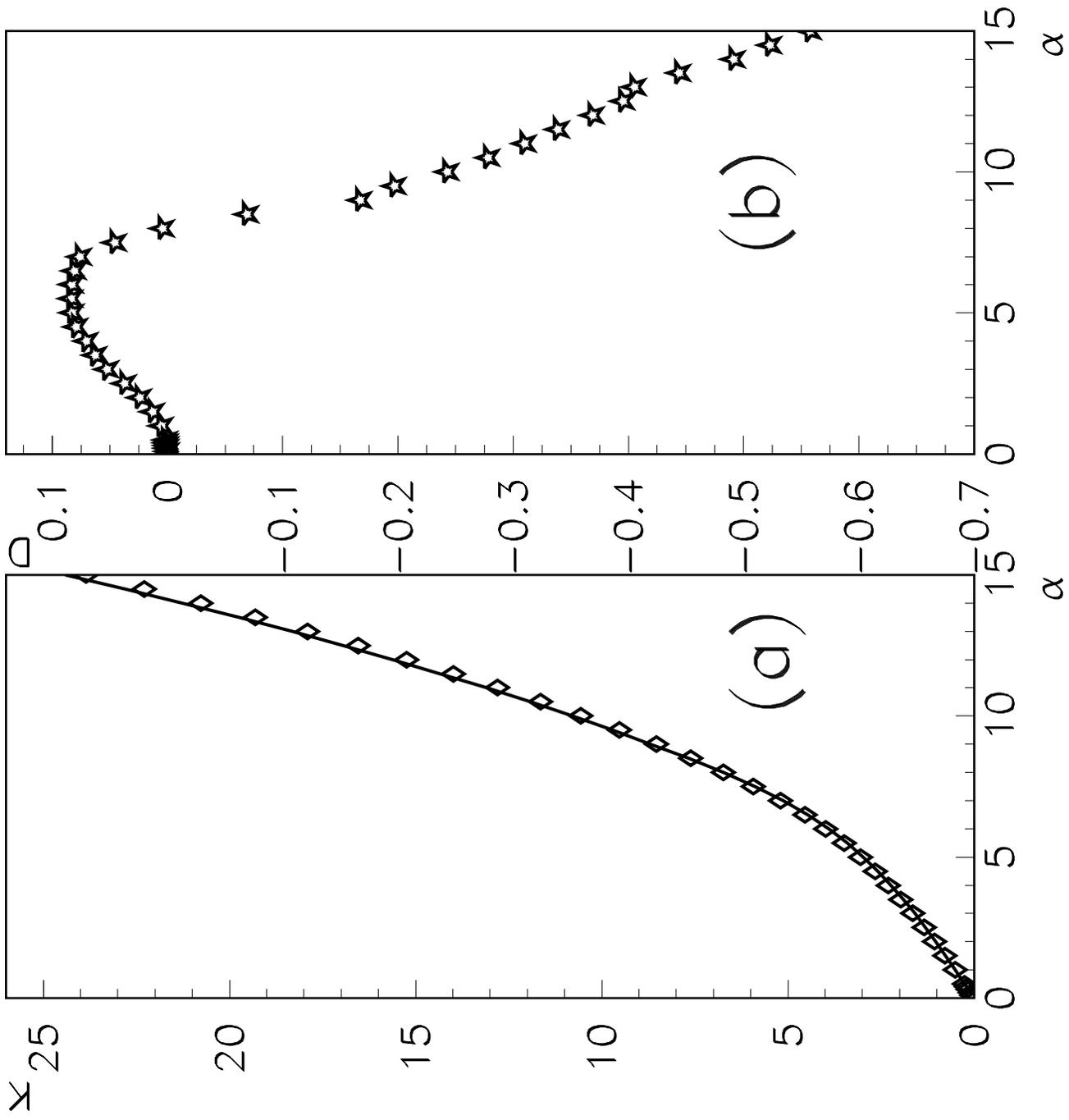
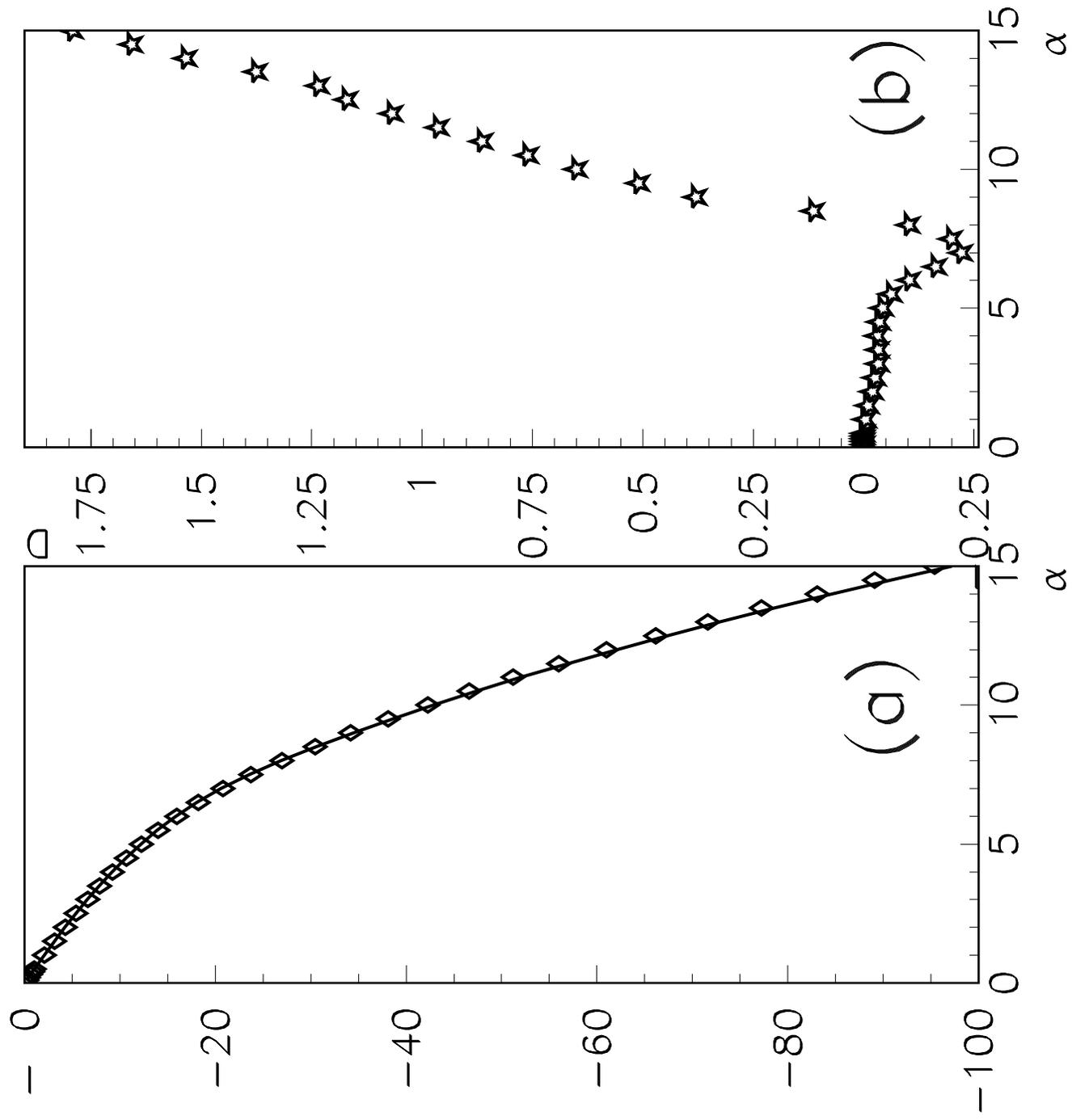
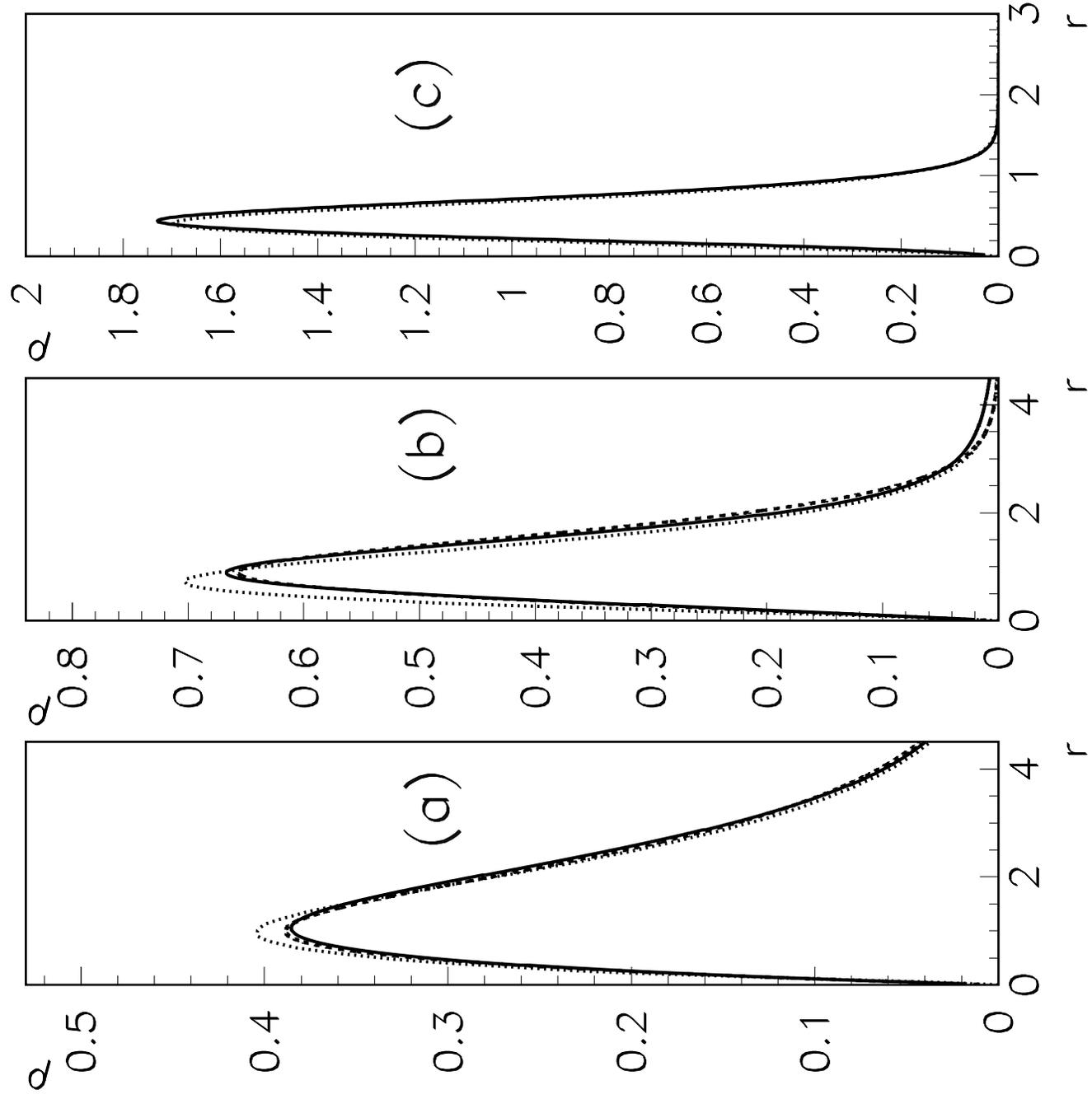
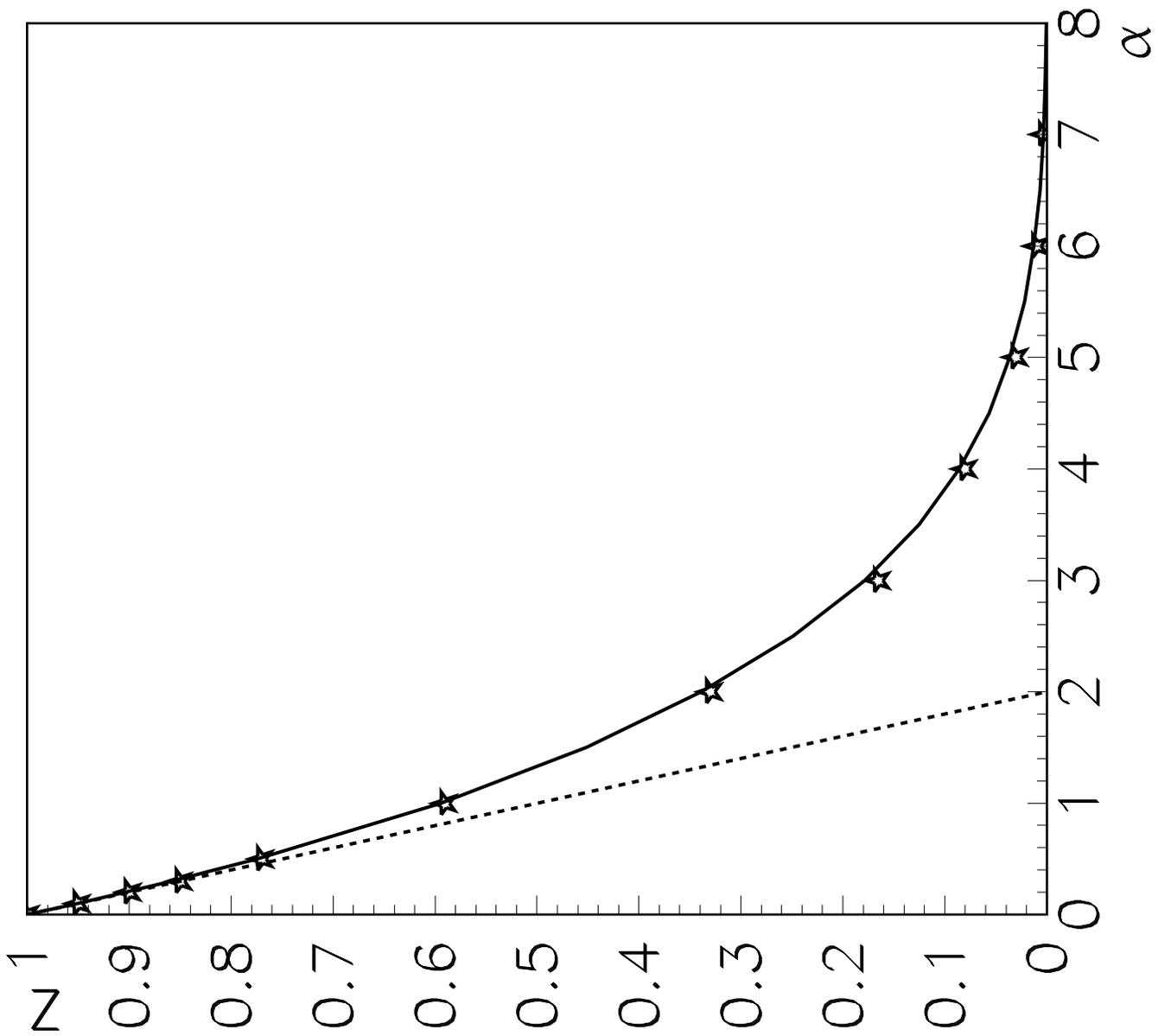
